\RequirePackage{fix-cm}
\documentclass[smallcondensed]{svjour3}     
\smartqed  
\usepackage{graphicx}
\usepackage[FIGTOPCAP]{subfigure} 
\usepackage{graphicx}
\usepackage{psfrag}
\usepackage{braket}
\usepackage{color}
\usepackage{amsmath}
 \usepackage{mathptmx}      
\usepackage{latexsym}
\usepackage{amssymb}
\newcommand{\Rey}{\mbox{Re}}  
 \journalname{under review}

\graphicspath{{./figures/}}
\begin{document}

\title{Relaminarization by steady modification of the streamwise velocity profile in a pipe
}


\author{J.\ K\"{u}hnen \and D.\ Scarselli \and M.\ Schaner \and B.\ Hof 
}


\institute{Jakob K\"{u}hnen \at
              Am Campus 1, A-3400 Klosterneuburg \\
              \email{jakob.kuehnen@ist.ac.at}           
}

\date{Received: date / Accepted: date}

\maketitle

\begin{abstract}
	
We show that a rather simple, steady modification of the streamwise velocity profile in a pipe can lead to a complete collapse of turbulence and the flow fully relaminarizes. Two different devices, a stationary obstacle (inset) and a device to inject additional fluid through an annular gap close to the wall, are used to control the flow. Both devices modify the streamwise velocity profile such that the flow in the center of the pipe is decelerated and the flow in the near wall region is accelerated. We present measurements with stereoscopic particle image velocimetry to investigate and capture the development of the relaminarizing flow downstream these devices and the specific circumstances responsible for relaminarization. We find total relaminarization up to Reynolds numbers of $6000$, where the pressure drop in the downstream distance is reduced by a factor of $3.4$ due to relaminarization. In a smooth straight pipe the flow remains completely laminar downstream of the control. Furthermore, we show that transient (temporary) relaminarization in a spatially confined region right downstream the devices occurs also at much higher Reynolds numbers, accompanied by a significant drag reduction. The underlying physical mechanism of relaminarization is attributed to a weakening of the near-wall turbulence production cycle. 
 
\end{abstract}

\section{Introduction}\label{sec:intro}

The control of turbulence to reduce skin friction is of great interest in a wide variety of technological applications, as frictional drag is a heavy consumer of energy and thus a possibly avoidable source of significant operating costs and massive carbon emissions. Numerous techniques for the control of turbulent flows have been proposed over the years. However, only a few methods of control have already been successfully implemented in practical engineering devices.

Flow control is commonly classified as either active or passive depending on whether or not external energy-input is required \cite{Gad-el-Hak00}. The most sophisticated and on a theoretical basis most elegant method is active feedback (closed loop) control of turbulence (see e.g. \cite{Moin1994,Lumley1998,Kim2007,Kasagi2009,Sharma2011,McKeon2013}). However, the practical implementation is technically highly demanding and expensive due to the required sensors for realtime flow measurements and elaborate actuators to control the flow. Predetermined (open loop) active techniques are usually characterized by greater simplicity and comparative ease of implementation. Among such techniques, modifying the flow through large-scale spanwise forcing created by boundary motion (wall oscillation or transverse traveling-wave excitation) or a body force has produced promising results (see e.g. \cite{Karniadakis2003,Quadrio2009,Auteri2010,Nakanishi2012,Tomiyama2013,Rabin2014}).

The wide range of passive control techniques comprises diverse approaches such as surface modifications by means of e.g.\ riblets \cite{Garcia2011}, grooves \cite{Frohnapfel2007}, shark-skin surfaces \cite{Dean2010}, hydrophobic walls \cite{Watanabe1999,Rothstein2010} or by forcing small optimal perturbations \cite{Fransson2006}. Furthermore, modifications of the fluid employing polymer additives \cite{White2008,Choueiri2017} and modifications of the flow field by means of honeycombs and screens \cite{Lumley1967,Laws1978}. Additionally, an appropriate streamline geometry obviously has a tremendous influence upon the structure of turbulent wall flows and can help to minimize drag \cite{Bushnell89}. 

Most of the above mentioned methods have in common that they reduce skin friction and decrease the turbulence level by some amount in the control area but can not totally extinguish turbulence and have no effect further downstream. However, the ultimate goal of turbulence control in terms of energy saving is relaminarization (also denoted as turbulent-laminar transition, reversion, retransition, and laminarization) of the flow, leaving aside the discussion on possible sublaminar friction in duct flows \cite{Bewley2009,Fukagata2009}.

Occasional evidence of relaminarization not determined by dissipation and the Reynolds number has been found when a turbulent flow is subjected to effects of acceleration, suction, blowing, magnetic fields, stratification, rotation, curvature and heating \cite{Sreenivasan1982}. For example, a peculiarity of curved pipes is that the threshold for the onset of subcritical turbulence is postponed and occurs at Reynolds numbers considerably larger than in straight pipes, causing low Reynolds number turbulent flow emerging from a straight pipe to relaminarize in a subsequent curved pipe \cite{Sreenivasan1983,Kuehnen2015}.

In particular, the effect of a favorable pressure gradient (FPG) on a boundary layer and the related case of accelerated pipe flow have received considerable attention concerning relaminarization. In accelerated pipe flow, i.e.\ during and subsequent to a rapid increase of the flow rate of an initially turbulent flow, the flow has been observed to transiently first visit a quasi-laminar state and then again undergo a process of transition that resembles the laminar-turbulent transition (see e.g. \cite{Lefebvre1989,Greenblatt1999,Greenblatt2004,He2013,He2015}). And a strong FPG imposed on a boundary layer has been found to have a damping effect on the growth of perturbations \cite{Corbett2000} and to cause a temporary state of relaminarization (see e.g. \cite{Patel1968,Blackwelder1972,Narasimha1973,Spalart1986,Warnack1998,Ichimiya1998,Mukund2006,Bourassa2009}). In experiments, the FPG is usually imposed on the flow by means of various types of convergence in wind tunnels. Closely related, e.g.\ \cite{Modi1997} successfully employ a 'moving surface boundary-layer control' to prevent or delay the separation of the boundary layer from the wall, where the moving surface is provided by rotating cylinders. Regarding pipe flow, \cite{Pennell1972} observed temporary relaminarization by fluid injected through a porous wall segment of the pipe. All these authors showed that a wall-bounded turbulent shear flow may relaminarize or, more accurately, transiently tend to a laminar-like or quasi-laminar state under certain suitable conditions, even if the Reynolds number is above criticality. 
However, in all cases inevitable retransition to turbulence was found at a later stage, rendering the use of the word relaminarization somehow misleading as it characterizes an intermediate but not the final result. Although various parameters have been proposed to quantify e.g.\ the acceleration level needed for relaminarization (for a compilation see \cite{Bourassa2009}) there is neither agreement on a precise criterion for the occurrence of relaminarization and how it can be triggered, nor on how its onset may be recognized.

Profound understanding of ways to control turbulence has emerged from studies elucidating the near-wall flow dynamics. Obviously the near-wall region is crucial to the dynamics of attached shear flows, as it is the region of the highest rate of turbulent energy production and of the maximum turbulent intensities. The dominant structures of the near-wall region are the streamwise velocity streaks and the quasi-streamwise vortices, and the dominant dynamics is a cyclic process characterized by the formation of velocity streaks from the advection of the mean profile by streamwise vortices, and the generation of vortices from the instability of the streaks, also referred to as self-sustaining process (see e.g. \cite{Hamilton1995,Waleffe1997,Jimenez2013,Brandt2014}). By clever use of 'wrong physics' for numerical experiments on modified turbulent channels \cite{Jimenez1999} showed that a local cycle of turbulence regeneration exists in the near wall region, which is independent of the outer flow. By numerically interrupting the cycle at various places they observed a breakdown of turbulence and eventual relaminarization. Also the aforementioned riblets are believed to directly weaken the quasi-streamwise vortices of turbulence regeneration, i.e.\ proposing that the drag reduction by riblets is due to the weakening of these vortices by the increase in spanwise friction at the wall \cite{Jimenez1994}.

\cite{Hof2010} observed an immediate collapse of single turbulent spots in the intermittent regime at relatively low Reynolds numbers. The breakdown of turbulence appeared if two turbulent spots were triggered too close to each other. They related the breakdown to a flattened streamwise velocity profile induced by the trailing spot, weakening the turbulence regeneration cycle beyond recovery. Their observation is also in qualitative agreement with predictions from a model explaining the emergence of fully turbulent flow in pipes and rectangular ducts \cite{Barkley2015}. From this model it was inferred that the excitability to the turbulent state only depends on the streamwise velocity component and hence an appropriate manipulation of the streamwise velocity profile may destroy the turbulent state.

Further investigations by \cite{Kuehnen2018} have recently shown that a modification of the velocity profile in a pipe by several means can lead to a complete collapse of turbulence and the flow can be forced to fully relaminarize also at higher Reynolds numbers. The annihilation of turbulence was achieved by a steady active open loop manipulation of the streamwise velocity component alone, greatly simplifying control efforts. In their numerical simulations \cite{Kuehnen2018} added an appropriate radially dependent body force term, $F(r)$, to the equation of motion, modifying the streamwise velocity profile to a more plug-like one. Furthermore, they presented four different experimental techniques to modify the velocity profile of turbulent flow, such that the resulting profile was (more) plug shaped and flat or even had velocity overshoots in the near wall region of the pipe. The first technique employed four rotors located inside the pipe to vigorously stir the flow. The second technique used wall-normal injection of additional fluid through 25 small holes placed consecutively in a helical fashion around the pipe. The third technique was by injecting fluid through an annular gap at the wall to accelerate the flow close to the wall. The fourth approach was by means of a movable pipe segment which was used to locally accelerate the flow at the wall. In all cases turbulence was shown to decay and the flow eventually relaminarized completely, up to Reynolds numbers of $100\,000$ in DNS and $40\,000$ in experiments. The experiments demonstrated that relaminarization occurred as a direct result of a particular velocity and shear-stress distribution especially in the wall region. The modified profile was shown to specifically suppress transient growth such that vortices do not efficiently create streaks.

Pipes and pipelines are central for the distribution of fluids throughout society, ranging from small diameter tubes in domestic settings to pipes in industrial plants and to large scale pipelines. The overall pumping costs surmount to billions of Euros per year and the frictional losses encountered in these flows are responsible for a significant part of the global energy consumption. Thus a huge amount of energy (pumping power due to frictional drag) could be saved if flows in pipes are laminar instead of turbulent. Pipe flow has a property that makes it particularly attractive for a way of turbulence control which totally extinguishes turbulence and establishes a laminar flow: the laminar state is stable to infinitesimal perturbations at all flow speeds \cite{Drazin1981}. Consequently, once relaminarization is achieved, the flow will remain laminar provided that the pipe is straight and smooth. Turbulence will only return if a sufficiently strong disturbance is encountered. It is therefore not necessary to apply relaminarization control techniques throughout the pipe but instead it suffices to implement control stations at fixed locations (for example behind bends) to ensure laminar flow in the straight downstream pipe sections.

In the present investigation we want to further explore the effect, scope and consequences of a modified streamwise velocity profile which relaminarizes the flow. In order to do so we insert an obstacle partially blocking the pipe (method 1) and inject fluid through an annular gap at the wall (method 2). Both approaches force the streamwise velocity profile in a similar manner. The inserted obstacle in a purely passive way, while the injection device can be considered steady open loop forcing. The obstacle device is better suited for measurements of the downstream flow field at various distances, as it can be easily mounted at any arbitrary position within the pipe. The device to inject fluid through an annular gap is better suited to continuously vary the amount of acceleration close to the wall and fathom out the maximum Reynolds number where relaminarization is possible. We present measurements with stereoscopic particle image velocimetry and investigate the development of the flow downstream these devices. Furthermore we try to elucidate the specific circumstances responsible for relaminarization.

The outline of the paper is as follows. In the next section we describe the experimental facility and the devices used to control the flow. Selected results of our extensive investigations and measurements are presented in section \ref{sec:results}. In section \ref{sec:discussion} we discuss the results and compare them to previous investigations concerned with relaminarization.


\section{Experimental facility}\label{sec:experimenalsetup}

The experimental setup consists of a basic pipe flow facility constituting fully turbulent flow in a straight long glass-pipe and two different kinds of interchangeable stationary flow management devices (FMDs) to control the flow. In the following section the facility and the different FMDs are described in detail.

\subsection{Facility}\label{subsec:facility}

\begin{figure} 
	\centerline{\includegraphics[clip,width=0.88\textwidth,angle=0]{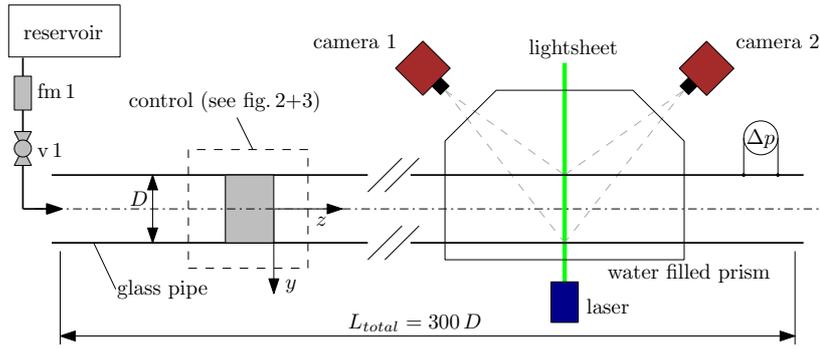}}
	\caption{\label{fig:sketch-setup_obst}Sketch of the experimental facility used to control turbulent pipe flow with two different kinds of devices. Stereoscopic PIV measurements of the flow field are made downstream of the devices (see figure \ref{fig:sketch-fmd} and \ref{fig:sketch-wi-nozzle} for details). The flow direction is from left to right. Drawing not to scale.}
\end{figure}

The setup consists mainly of a glass-pipe with inner diameter $D=30\pm0.01$\,mm and a total length of 9\,m ($300\,D$) made of 1 meter sections. Figure \ref{fig:sketch-setup_obst} shows a sketch of the facility. Gravity driven water enters the pipe from a reservoir. The flow rate $Q_m$ is measured by means of an electromagnetic flowmeter (fm\,1) in the supply pipe. The Reynolds number ($\Rey= U D/{\nu}$, where $U$ is the mean velocity and $\nu$ the kinematic viscosity  of the fluid) can be adjusted by means of a valve (v\,1) in the supply pipe.

The velocity vector field is measured $\sim250\,D$ downstream from the inlet at the position of the lightsheet. The measurement plane is perpendicular to the streamwise flow direction (pipe $z-$axis). All three velocity components within the plane are recorded using a high-speed stereo PIV system (Lavision GmbH) consisting of a continuous laser (Fingco 532H-2W) and two Phantom V10 high-speed cameras with a full resolution of $2400\times 1900$\,px. Spherical glass particles (sphericel, mean diameter 13\,$\mu$m, Potter Industries) are used to seed the flow. Around the measurement plane the pipe is encased by a water filled prism such that the optical axes of the cameras are perpendicular to the air--water interface to reduce refraction and distortion of the images.

A differential pressure sensor (DP\,103, Validyne) is used to measure the pressure drop $\Delta p$ between two pressure taps. It is used for a straightforward detection of the flow state based on the large difference between turbulent and laminar friction factors ($f_T$ and $f_L$). For this purpose the sensor is placed downstream the FMD as indicated in fig.\ \ref{fig:sketch-setup_obst}, and the pressure taps are separated by $30\,D$ in the streamwise direction.

For reference a non-dimensional Cartesian coordinate-system ($x,y,z$) = ($\tilde{x}/D$, $\tilde{y}/D$, $\tilde{z}/D$) is used as indicated in figure \ref{fig:sketch-setup_obst}. The origin of the coordinate system is always located at the downstream end of the FMD. The respective Cartesian velocity components ($\tilde{u},\tilde{v},\tilde{w}$) are made non-dimensional either with the mean velocity $U$, yielding ($u,v,w$), or with the friction velocity $w_{\tau}$, yielding $y^{+}$ and $w^{+}$. For $w_{\tau}$ we use the estimate $w_{\tau}=U (f/8)^{0.5}$ \cite{Pope2001}, where the friction factor $f$ is obtained from pressure drop measurements. Velocity fluctuations are denoted by $'$, $<>$ denotes cross sectional averaging and a bar averaging over time.

\subsection{Orifice plate obstacle (method 1)}\label{subsec:orifice}

\begin{figure}
	\centerline{\includegraphics[clip,width=0.86\textwidth,angle=0]{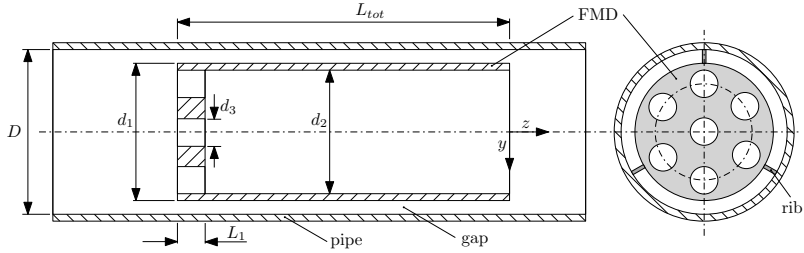}}
	\caption{\label{fig:sketch-fmd}Sketch (sideview and frontview) of the orifice plate obstacle (obstacle-FMD). The device can be mounted at any axial position within the pipe (the ribs provide interference fit). The flow direction is from left to right. Drawing not to scale.}
\end{figure}

The first device to control the flow is a stationary obstacle or inset which can be mounted within the pipe. Figure \ref{fig:sketch-fmd} shows a sketch of the device. It is made of a custom made thin-walled tube of $L_{tot}=200$\,mm total length with an outer diameter of $d_1=28$\,mm and an inner diameter of $d_2=26$\,mm. To facilitate flow visualization the tube is made of plexiglass. The tube is closed at the upstream end with a perforated plate with a wall thickness of $L_1=5$\,mm. 7 holes with diameter $d_3=3.3$\,mm are drilled into that plate. The device can be mounted concentrically within the pipe at any axial position by means of three small streamwise ribs (interference fit).

As a result 78.6\% of the pipe are blocked at the upstream end of the device. The flow is divided into two separate parts. One part of the flow goes through the 1\,mm annular gap along the pipe wall, while the other part of the flow goes through the perforated plate in the bulk. The device thus forms a deliberate obstacle or obstruction acting as a spatially extended volume forcing on the flow. Its sole purpose is to tailor the velocity distribution of the flow at the downstream end of the device (in the plane $z=0$) in a controlled way. For reference in the following text the device is referred to as obstacle-FMD.

Devices with several different blockage ratios, gap widths, more or less holes with different diameters have also been tested. Although there is certainly still room for further optimization, it was found that with reasonable effort the specific dimensions mentioned above seem to work best in terms of relaminarization capability. The device is quite sensitive to changes. E.g., already if the diameter $d_3$ of the holes is $\leq2.8$\,mm or $\geq4$\,mm instead of 3.3\,mm and all other dimensions are left unchanged, the device is not able to relaminarize the flow at any Reynolds number.

\subsection{Annular gap injection nozzle (method 2)}\label{subsec:nozzle}

\begin{figure}
	\centerline{\includegraphics[clip,width=0.8\textwidth,angle=0]{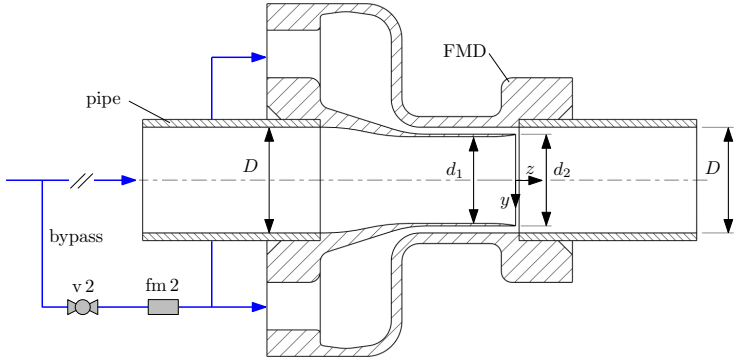}}
	\caption{\label{fig:sketch-wi-nozzle}Sketch of the control device acting as injection nozzle through an annular gap (1\,mm-FMD and 2\,mm-FMD). Fluid is taken from the main pipe via a bypass and then re-injected through a concentric gap close to the wall. The device is mounted within two pipe segments. The bypass is equipped with a valve (v\,2) and a flowmeter (fm\,2). The flow direction is from left to right. Drawing not to scale.}
\end{figure}

To complement and extend the purely passive control mechanism of the obstacle-FMD we designed a device which allows to inject fluid into the main pipe through a small annular concentric gap close to the pipe wall as shown in figure \ref{fig:sketch-wi-nozzle}. The modification of the velocity profile resulting from this active open-loop control of the flow is similar to the stationary obstacle, but the amount of injected fluid and hence the level of acceleration close to the pipe wall can be controlled and continuously adjusted via a valve in the feeding line. In the reported setup the feeding line is implemented as a bypass that takes fluid from upstream the main pipe and re-injects it through the device. However, the injection device could just as well be used without the bypass by injecting fluid from an external reservoir via a pump. This method was found to be equally effective and relaminarization up to similar $\Rey$ values could be achieved.

We tested two different devices by systematically varying the injected flow rate at different Reynolds numbers, one device with an annular gap of 1\,mm and one with an annular gap of 2\,mm. In the following text these devices are referred to as 1\,mm-FMD and 2\,mm-FMD.

As sketched in figure \ref{fig:sketch-wi-nozzle} the main pipe is slightly narrowed in a short range just upstream the injection point (1\,mm-FMD: $d_1=26.6$\,mm, $d_2=28$\,mm, open gap area $A_1=91.1$\,mm$^{2}$; 2\,mm-FMD: $d_1=24.6$\,mm, $d_2=26$\,mm, open gap area $A_2=175.9$\,mm$^{2}$). At a small backward facing step ($z=0$, the plane of confluence) the fluid coming from the bypass is axially injected into the main pipe through the annular gap close to the wall. Based on the measured bypass flow rate $Q_{bp}$ and the open gap area the mean velocity $U_{inj}$ of the injected flow in the plane of confluence can be calculated. As the total flow rate $Q_m$ is measured in the supply pipe (see section \ref{subsec:facility}), the specified Reynolds number in the main pipe applies to the range upstream the bypass and downstream the confluence at $z=0$. The Reynolds number $\Rey_n$ and the mean velocity $U_{n}$ in the slightly narrowed part of the device is calculated based on $d_1$ and $Q_m-Q_{bp}$.

In all measurements reported the flow rate was chosen such that $\Rey_n\gtrsim3000$ to ensure fully turbulent flow in the slightly narrowed section (as for $\Rey_n\lesssim2800$ the main flow may become (intermittently) laminar already without any forcing \cite{Avila2013a}). For the 2\,mm-FMD this criterion was met for $\Rey\gtrsim4000$. 

\section{Results}\label{sec:results}

In this section we present the results of the measurements split into two subsections for each kind of device. Representative sets of stereoscopic PIV-measurements are shown for Reynolds numbers where the flow relaminarizes and where it does not. The FMDs were first investigated at different Reynolds numbers by means of mere visualization. Neutrally buoyant anisotropic particles \cite{Matisse1984} were added as tracer particles and the flow in the pipe was illuminated by means of LED string lights along the whole length of the pipe to be able to observe the development of the flow field downstream of the FMDs. The observations were also recorded with a video camera. A selected prime example of a relaminarizing flow is made available in the online supporting material.

From visual observations in combination with pressure drop measurements we obtained very reliable information whether or not the flow completely relaminarized at a certain Reynolds number. Although relaminarization is a gradual process \cite{Sreenivasan1982} and the flow tends only asymptotically to the fully developed laminar parabolic Hagen-Poiseuille profile \cite{Durst2005}, it is accompanied by drastic changes in the structure and dynamical behaviour of the flow. The mean velocity profile departs from the well-known law of the wall, the friction factor exhibits a substantial decrease and Reynolds stresses become negligible, the turbulence intensity goes down, etc. Visually most striking and conspicuous is the total decay of any eddying motion, as all velocities normal to the streamwise direction fade away.

More difficult to capture and quantify is the transient state of ongoing relaminarization, also referred to as laminarescent. It signifies the earlier stages of relaminarization - loosely, a precursor to relaminarization - in which large departures occur from the turbulent state \cite{Sreenivasan1982}. However, remaining perturbations in the flow may trigger retransition to turbulence. It is only when the flow has reached a sufficiently laminar state that it remains laminar for the remainder of the pipe. As will be shown further down, monitoring the flow at $z\approx100-150$ provides a clear and unambiguous indication whether or not the flow has relaminarized completely.

\subsection{Measurements of uncontrolled turbulent flow for reference}\label{subsec:results-reference}

\begin{figure}
	\centerline{\includegraphics[clip,width=0.68\textwidth,angle=0]{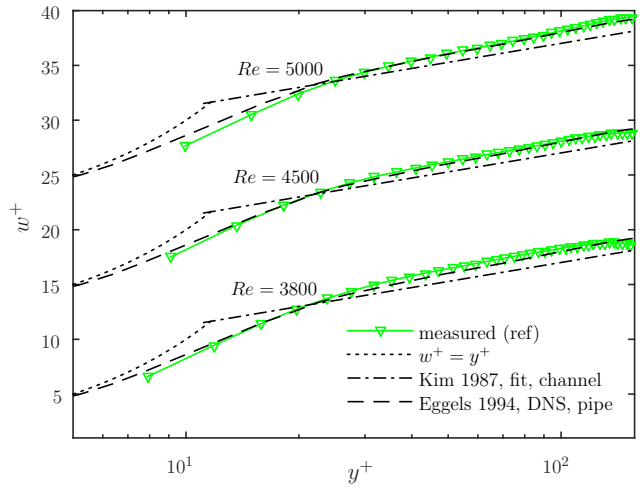}}
	\caption{\label{validation-velprofiles}A comparison of the measured uncontrolled turbulent flow (ref) with the the law of the wall (fit from \cite{Kim1987}, $w^{+}=2.5 \ln y^{+}+5.5$) and DNS in a pipe from \cite{Eggels1994}. The viscous sublayer is indicated by $w^{+}=y^{+}$.}
\end{figure}

\begin{figure}
	\centerline{\includegraphics[clip,width=0.5\textwidth,angle=0]{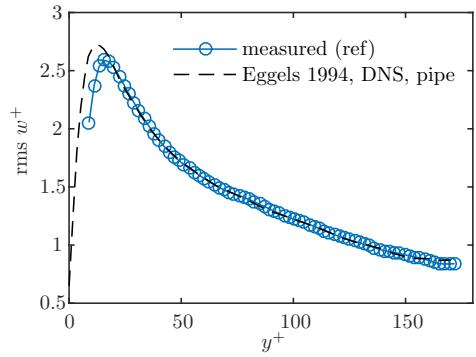}}
	\caption{\label{validation-rms-4500}Streamwise rms profile measured at $\Rey=5000$ scaled on inner variables. The dashed line shows the results from \cite{Eggels1994} for comparison.}
\end{figure}

To be able to compare the controlled, relaminarizing flow to uncontrolled fully turbulent flow we first measured the flow without FMDs. The respective measurements are going to be used for reference in the remainder of the present work. The velocity profiles measured at $\Rey=3800$, $\Rey=4500$ and $\Rey=5000$ normalized by the inner-layer variables are plotted in figure \ref{validation-velprofiles} in the usual semi-logarithmic manner. For comparison, the law of the wall is plotted with the K\'{a}rm\'{a}n constant $k=0.4$ and the additive constant $A=5.5$ (to compensate the low-Reynolds number effect, see \cite{Kim1987}). Furthermore, DNS results by \cite{Eggels1994} for pipe flow at $\Rey=5300$ are shown. \cite{Eggels1994} demonstrated that the mean velocity profile in a pipe fails to conform to the law of the wall, in contrast to channel flow. This is clearly reaffirmed by our measurements. While for $\Rey=3800$ our measured velocity profile is slightly above the DNS data of \cite{Eggels1994}, their data coincide very well with our results for $\Rey=4500$ and $\Rey=5000$. \cite{Eggels1994} stated a best-fit of $k=0.35$ with an additive constant $A=4.8$. For $\Rey=5000$ we find a best-fit of $k=0.33$ and $A=4.2$ for $40<y^{+}<120$ or $k=0.34$ and $A=4.6$ for $40<y^{+}<100$. The near wall region is only partly resolved in the measured velocity profiles. The smallest values of $y^{+}$ are 7.9, 9.1 and 10 respectively, corresponding to $\sim0.08$\,mm.

The root-mean-square (rms) values of the streamwise fluctuating velocity, normalized by the friction velocity, are plotted in figure \ref{validation-rms-4500} for $\Rey=5000$. Obviously, the major part of the fluctuations occurs close to the wall. For comparison, DNS results by \cite{Eggels1994} for pipe flow at $\Rey=5300$ are shown. The peak value (2.6) of the measured rms is in excellent accordance and the position of the peak ($y^{+}=18.2$) as well as the overall trend of the data is in good accordance with previous investigations (see also \cite{Mochizuki1996}).

\subsection{Orifice plate obstacle (inset)}\label{subsec:results-orifice}

\begin{figure}
	\centerline{\includegraphics[clip,width=0.96\textwidth,angle=0]{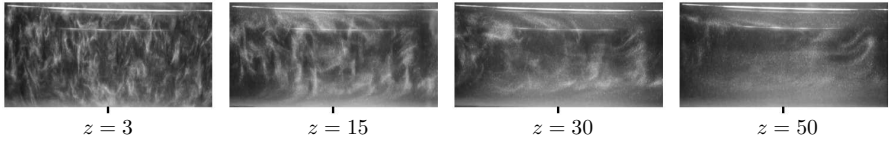}}
	\caption{\label{fig:relam-evo-pics}Still pictures from supplementary movie 1 (see online materials) of a short section of the pipe at $\Rey=4000$. The camera was following the flow downstream the obstacle-FMD at the mean velocity $U$. The difference between turbulent flow (at $z=3$) and laminar flow (at $z=50$) is unambiguous.}
\end{figure}

For the visual observation of the flow field we placed the obstacle-FMD $80\,D$ downstream from the entrance of the pipe where the turbulent flow can be considered fully developed. We varied the Reynolds number between 2500 and 5000 and found a continuously laminar flow field at $z=150$ up to $\Rey\approx4200$. At slightly higher Reynolds numbers we observed laminar-turbulent intermittency. For $\Rey\gtrsim4400$ we found only turbulent flow at $z=150$. Inspection of the flow field in the closer downstream vicinity ($0< z\lesssim10$) of the obstacle-FMD shows a rather turbulent zone right downstream the device where strong cross-stream motion is obviously present. But for $\Rey\lesssim4200$ all visible perturbations in the flow field decay until the flow appears clearly laminar at $z\approx50$. This evolution is well observable in figure \ref{fig:relam-evo-pics}, depicting still pictures from supplementary movie 1 (see online materials; see also \cite{Kuehnen2015video}), where the camera follows the flow at approximately the mean velocity on its journey from upstream of the device until it has relaminarized downstream. A laminarescent part of the flow as in figure \ref{fig:relam-evo-pics} at $z=15$, where the turbulence intensity is obviously reduced by some amount in the downstream vicinity of the device, could be observed up to $\Rey\approx4800$.

\begin{figure}
	\begin{center}
		\subfigure[$\Rey=3800$]{\includegraphics[clip,width=0.88\textwidth]{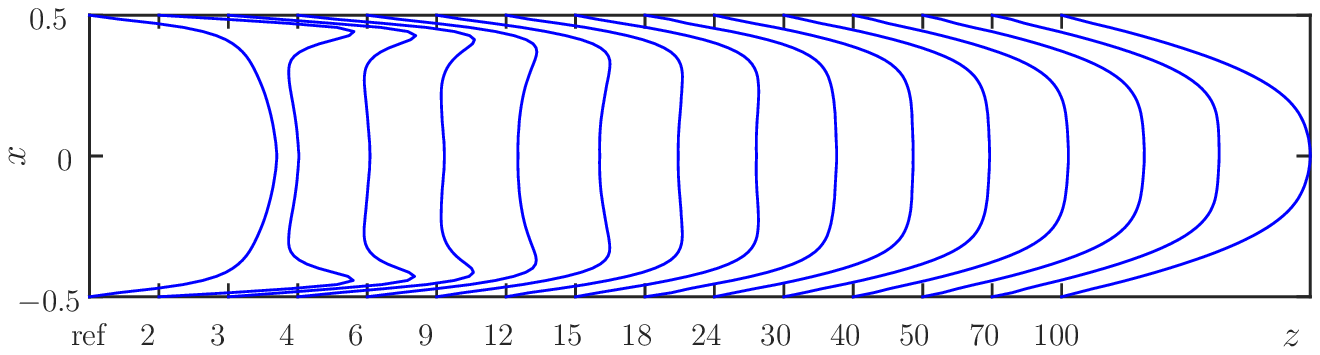}}\qquad
		\subfigure[$\Rey=4500$]{\includegraphics[clip,width=0.88\textwidth]{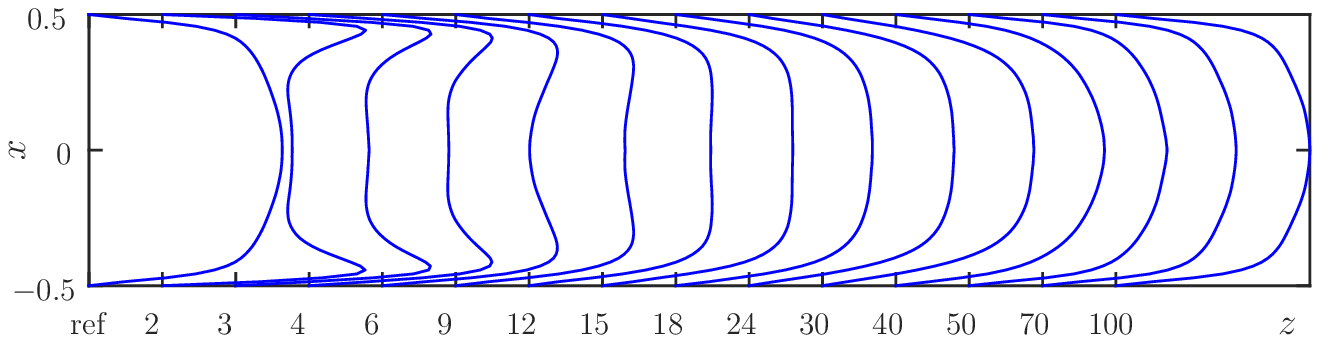}}\qquad
		\subfigure[centerline velocities]{\includegraphics[clip,width=0.96\textwidth]{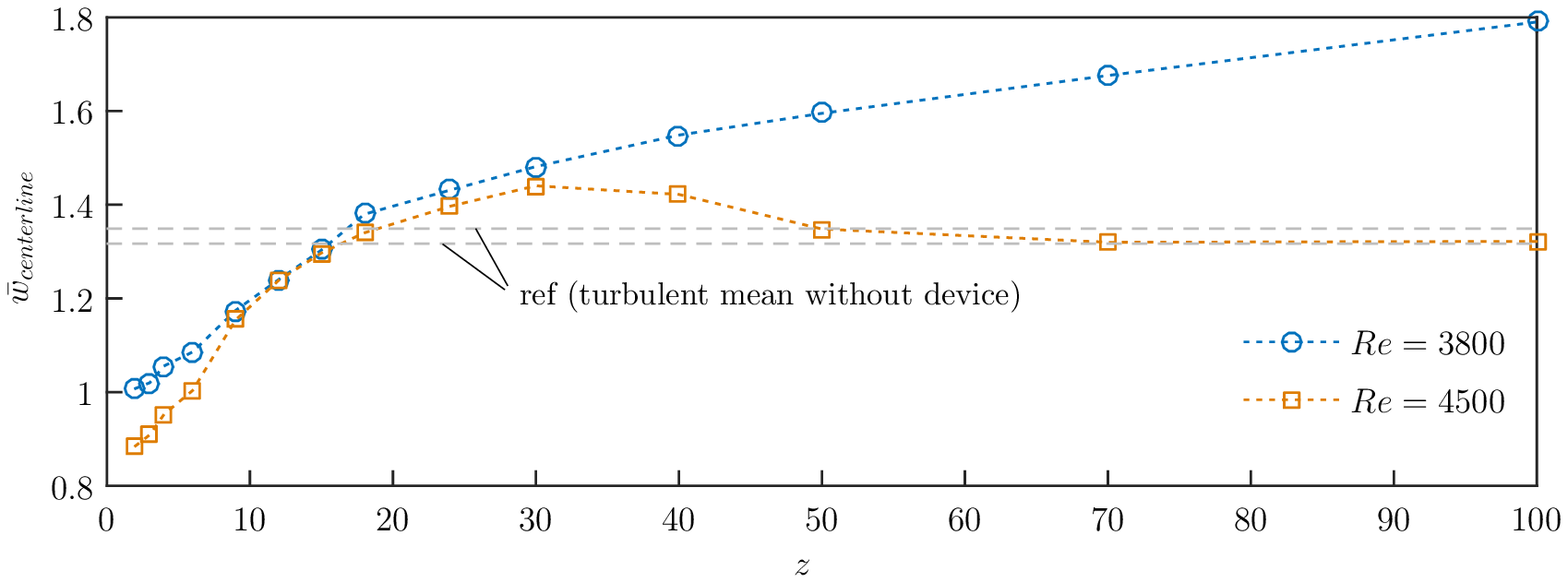}}\\
	\end{center}
	\caption{\label{fig:Orifice7holes_profile_evo}Downstream evolution of mean streamwise velocity profiles $(\bar w)$ in the plane $y=0$ measured at (a) $\Rey=3800$ and (b) $\Rey=4500$. In each plot the profile of uncontrolled turbulent flow is shown for reference (ref) at the left. (c) Respective levels of the axially developing centerline velocity.}
\end{figure}

To characterize the flow downstream the device we took PIV-measurements for two representative cases at $\Rey=3800$ and $\Rey=4500$ and 14 axial stations downstream the FMD: at $z=2,\,3,\,4,\,6,\,9,\,12, 15, 18, 24, 30, 40, 50, 70$ and 100. The development of the streamwise velocity $\bar w$ in the plane $y=0$ is shown in figure \ref{fig:Orifice7holes_profile_evo} (a) and (b). Each profile is calculated from 200 independent vector fields and azimuthally and temporally averaged. For reference, the most left profile is showing the respective measurement of uncontrolled turbulent flow (ref). Figure \ref{fig:Orifice7holes_profile_evo} (c) provides the respective levels of the centerline velocity for quantification. As can be seen from the figure, at $\Rey=3800$ the flow fully relaminarizes, while at $\Rey=4500$ retransition to turbulence takes place.

A particularly noteworthy effect of the device on the time-averaged velocity profiles is the increase in the velocity gradient at the wall. At $z=2$ the profiles at both Reynolds numbers exhibit characteristic overshoots of fast fluid close to the wall (peak at $x\pm0.44$) and a clearly decreased centerline velocity ($\sim1$ and $\sim0.9$ respectively). The bulk area depicts almost a plateau, in which the axial velocity is approximately constant, stretching from $-0.32\lesssim x \lesssim0.32$. The plateau is even more flat at $z=3$ and 4, where the peak of the velocity overshoot has moved slightly towards the center ($x\pm0.43$ and $x\pm0.41$ respectively). The increase in the velocity gradient close to the wall, however, does not persist for more than a few pipe diameters. At $z\approx15$ the two peaks have disappeared completely and the plateau covers the range $-0.25\lesssim x \lesssim0.25$. Simultaneously, the initially steep gradient in the axial velocity close to the wall has decreased considerably. 
Concerning the centerline velocity the value of uncontrolled turbulent flow is regained around $z=15-18$. Interestingly, for both Reynolds numbers $\bar w_c$ rises clearly above the uncontrolled level in the subsequent range $18\lesssim z\lesssim50$. 

The completely different downstream behavior of the flow at $\Rey=3800$ and $\Rey=4500$ becomes apparent only for $z\gtrsim30$. The flow at $\Rey=3800$ keeps developing towards a laminar parabolic profile. At $z=100$ the centerline velocity has reached a value of $\sim1.8$. At this Reynolds number it takes yet another $\sim120\,D$ \cite{Durst2005} to reach the fully developed state under perfect conditions (the laminar profile is sensitive to small disturbances, e.g.\ due to thermal convection or minor misalignments of the pipe segments \cite{Doorne07}). But the centerline velocity of the flow at $\Rey=4500$ keeps increasing only till $z\approx30$. The mean turbulent centerline velocity is clearly surpassed at $z\gtrsim20$. However, in the range $30\lesssim z\lesssim40$ the centerline velocity starts to fall back towards the turbulent mean.

\begin{figure}
	\begin{center}
		\subfigure[turbulence intensity (streamwise)]{\includegraphics[clip,width=0.96\textwidth]{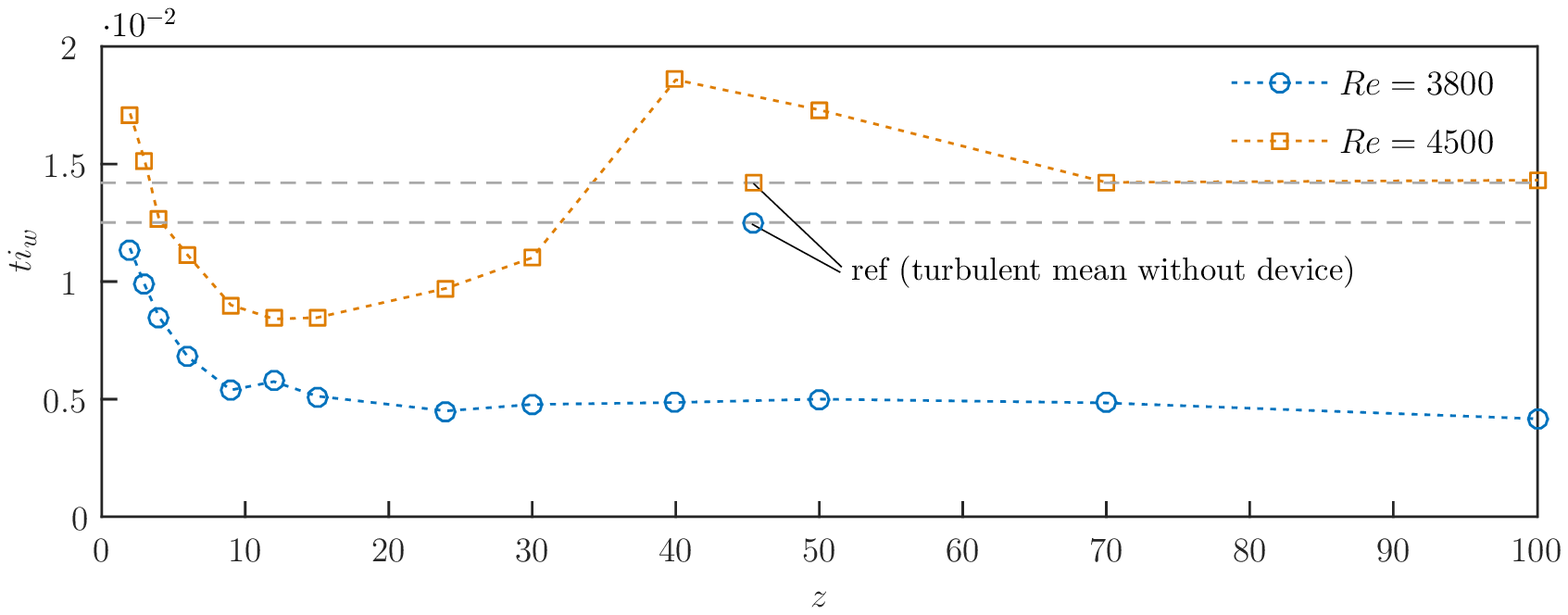}}\qquad
		\subfigure[turbulence intensity (inplane)]{\includegraphics[clip,width=0.96\textwidth]{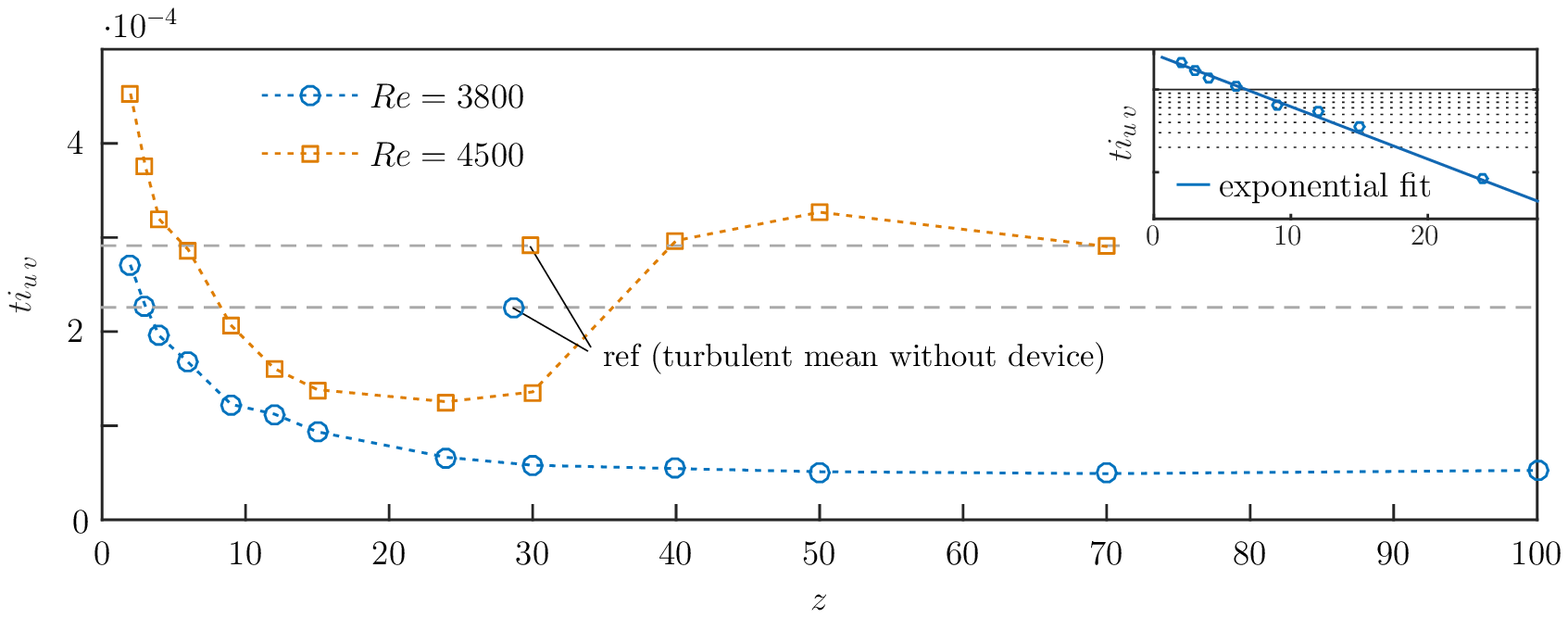}}\\
	\end{center}
	\caption{\label{fig:Orifice7holes_ti}Downstream evolution of the streamwise ($ti_{w}$) and inplane ($ti_{u,v}$) turbulence intensity. The respective level of uncontrolled turbulent flow is shown for reference (ref). The inset in (b) displays an exponential fit to the data at $\Rey=3800$ for $z=2-30$, plotted on a log scale. }
\end{figure}

The scenario of complete relaminarization at $\Rey=3800$ and of final retransition to turbulence at $\Rey=4500$ is also apparent from the axial development of the streamwise and inplane turbulence intensity ($ti_{w}$ and $ti_{u,v}$) as plotted in figure \ref{fig:Orifice7holes_ti} (a) and (b), where
\begin{equation}\label{w_ti}
ti_{w} =<(\sqrt{\bar{(w')^2}})/U>\,=\,<\dfrac{w_{rms}}{U}>
\end{equation} and
\begin{equation}\label{uv_ti}
ti_{u,v} =<(\sqrt{\bar{(u')^2}}+\sqrt{\bar{(v')^2}})/2U>\,=\,<\dfrac{u_{rms}+v_{rms}}{2U}>.
\end{equation}
Right downstream the device the levels of turbulence intensity are slightly increased above the turbulent reference flow. It should be noted however that in the bulk region of the cross-section right downstream of the obstacle-FMD remaining flow perturbations induced by the perforated plate at the upstream end of the obstacle-FMD may still be significant, as $L_{tot}=200$\,mm (see figure \ref{fig:sketch-fmd}) is not sufficiently long to regain streamwise invariance (see e.g. \cite{Barbin1963,Doherty2007}). The increase in turbulence intensity can hence not be clearly attributed to the accelerated part of the flow coming through the gap.

Already at $z\gtrsim3$ ($\Rey=4500$) $ti_{w}$ drops below the level of uncontrolled turbulent flow. The same applies regarding $ti_{uv}$ at $z=4$ ($\Rey=3800$) and $z=6$ ($\Rey=4500$) respectively. While for $\Rey=3800$ $ti_{uv}$ decreases exponentially until $z\sim30$ where the stable level of unavoidable measurement noise is reached, a qualitatively similar decrease in $ti_{uv}$ at $\Rey=4500$ is perceived  only until $z=15$. For $15\lesssim z\lesssim30$ $ti_{uv}$ is almost constant, followed by a steep increase which even leads to an overshoot above the mean turbulent level for $40\lesssim z<70$. Further downstream the flow seems to resemble an uncontrolled turbulent flow again. $ti_{w}$ exhibits a qualitatively very similar development, but the increase back to the turbulent level at $\Rey=4500$ starts a little earlier (around $z=15$) and the overshoot above the turbulent level takes place more rapidly and pronouncedly ($+30\%$ at $z=40$). At $\Rey=3800$ $ti_{w}$ seems to reach the laminar (noise) level in the range $z=10-15$. The remaining velocity fluctuations can be attributed to measurement noise.

\begin{figure}
	\begin{center}
		\subfigure[$\Rey=3800$]{\includegraphics[clip,width=0.7\textwidth]{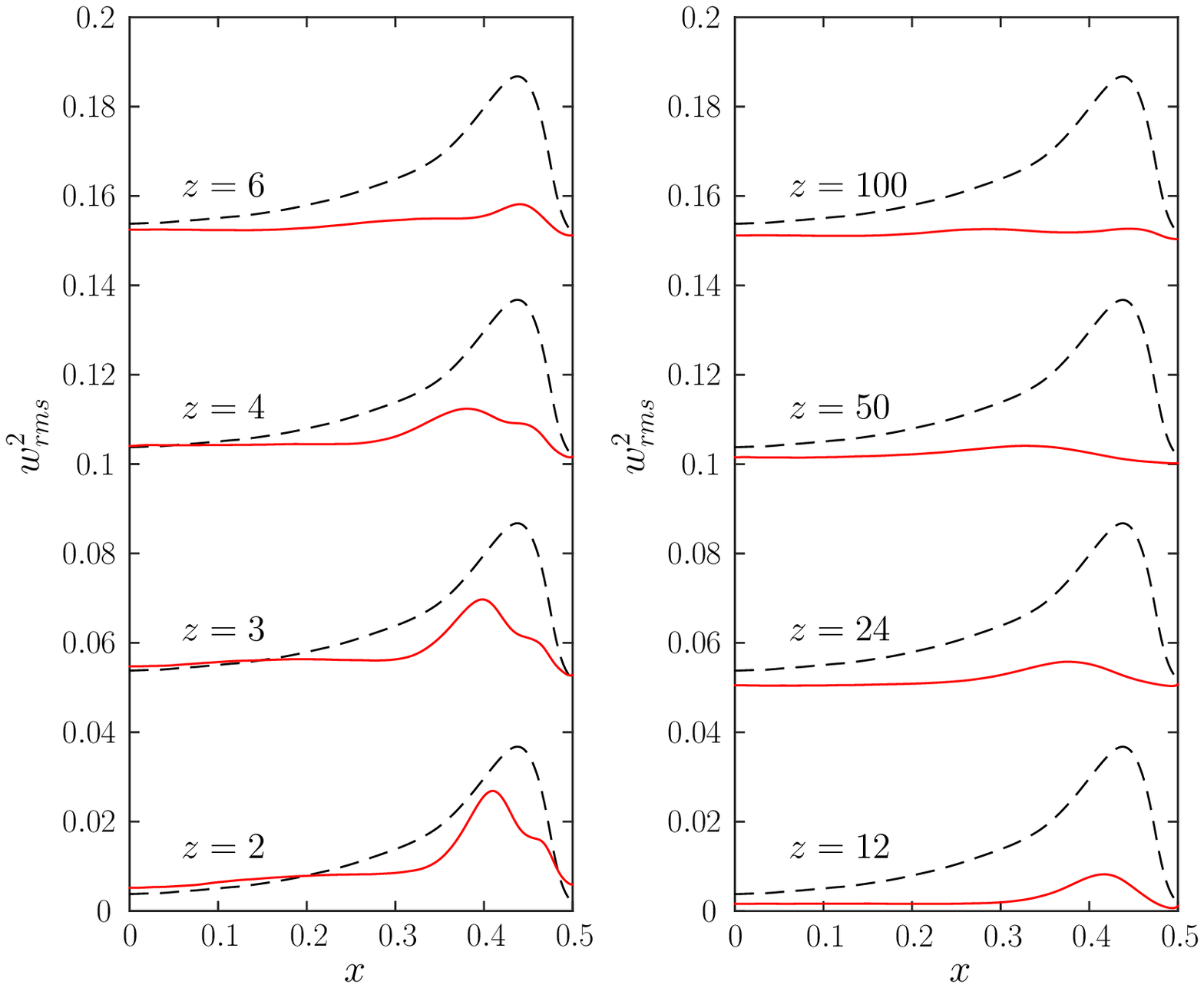}}\qquad
		\subfigure[$\Rey=4500$]{\includegraphics[clip,width=0.7\textwidth]{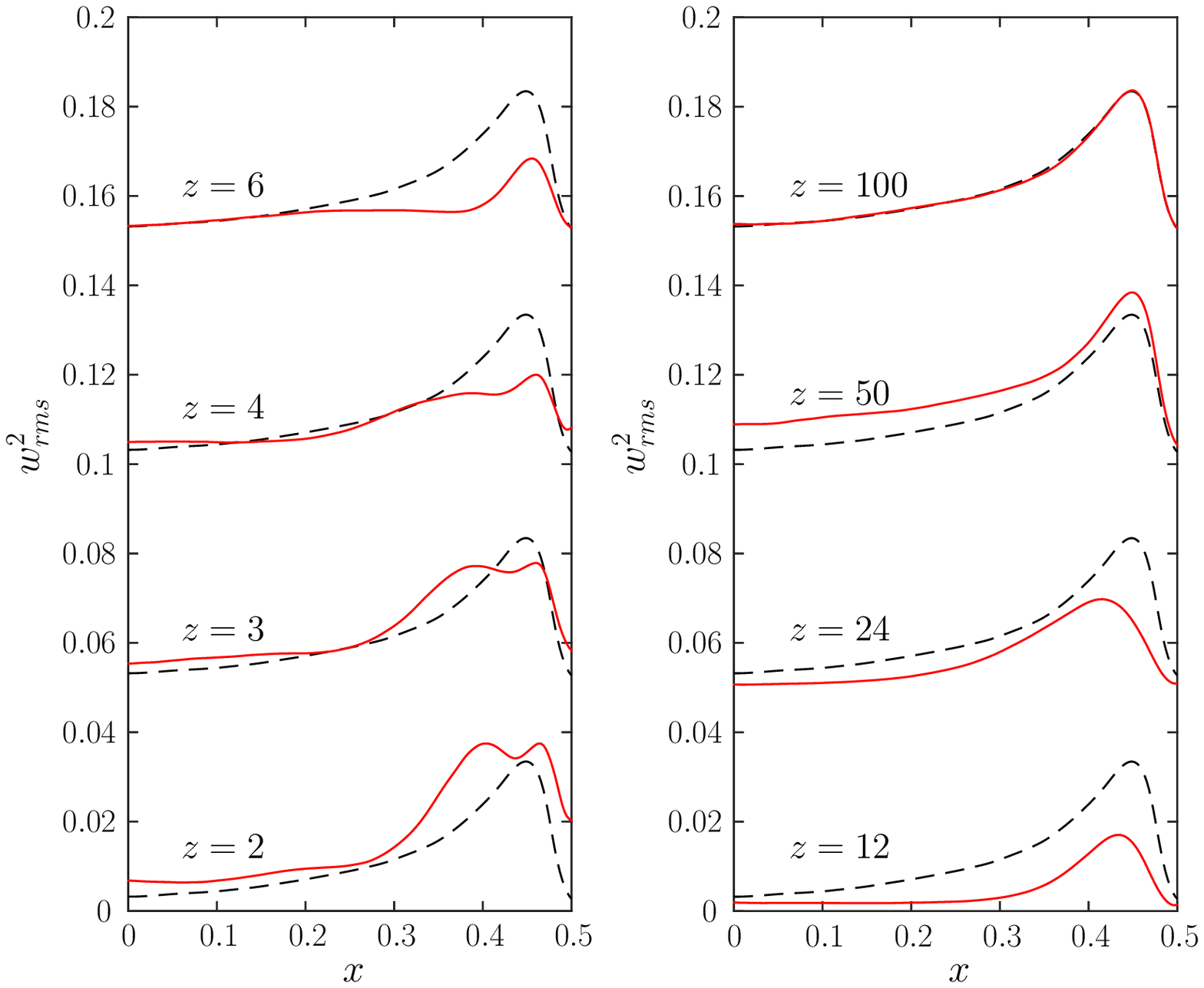}}\\
	\end{center}
	\caption{\label{fig:Orifice7holes_w_rms_evo_SUB}Downstream evolution of the streamwise velocity fluctuations ($w_{rms}^2$) at 8 selected axial stations for $\Rey=3800$ and $\Rey=4500$. The dashed line is showing uncontrolled turbulent flow for reference.}
\end{figure}

To elucidate the downstream development of turbulence fluctuations within the cross-section of the pipe, the streamwise velocity fluctuations ($w_{rms}^2$) at selected axial stations are depicted in figure \ref{fig:Orifice7holes_w_rms_evo_SUB}. The reference measurement of uncontrolled turbulent flow (see also fig.\ \ref{validation-rms-4500}) at the respective Reynolds number is plotted as a dashed line.

In the core region of the cross-section right downstream of the obstacle-FMD the flow is only slightly changed for both Reynolds numbers compared to the uncontrolled flow, recognizable by the increased rms level for $x\lesssim0.2$ at $z=2$. This is apparently due to the perturbations caused by the perforated plate at the upstream end of the obstacle-FMD. Closer to the wall a more prominent deviation from the uncontrolled flow is visible, depicting also a clear cut qualitative difference between the measurements at $\Rey=3800$ and $\Rey=4500$. For $\Rey=3800$ the maximum of streamwise fluctuations close to the wall has decreased below the level of uncontrolled flow already at $z=2$. Additionally, the maximum has significantly moved towards the centerline (apparent until $z=4$). As $z$ increases there is a substantial reduction in turbulence level within the whole cross-section. Initially the turbulent fluctuations decay very quickly close to the wall and more slowly in the bulk region. For $z>30$ the flow can be considered laminar and the overall turbulence intensity does not change significantly anymore (see also figure \ref{fig:Orifice7holes_ti}).

For $\Rey=4500$ on the other hand the maximum of streamwise fluctuations at $z=2$ has not dropped significantly and is comparable to uncontrolled turbulence. A double-humped, much thickened shape of the near-wall peak can be observed instead (visible up to $z=4$). However, for $z=4-24$ the peak close to the wall clearly decreases beyond the level of uncontrolled flow and the overall turbulence intensity is substantially reduced. Yet, at $z=50$ the near wall region looks similar to the reference flow again. In the core region the rms level is even increased beyond the reference flow, yielding the increased level in turbulence intensity as depicted in figure \ref{fig:Orifice7holes_ti}. Finally, at $z=100$, the flow has completely returned to the uncontrolled level.

\subsection{Annular gap injection nozzle}\label{subsec:results-injection}

\begin{figure}
	\centerline{\includegraphics[clip,width=0.98\textwidth,angle=0]{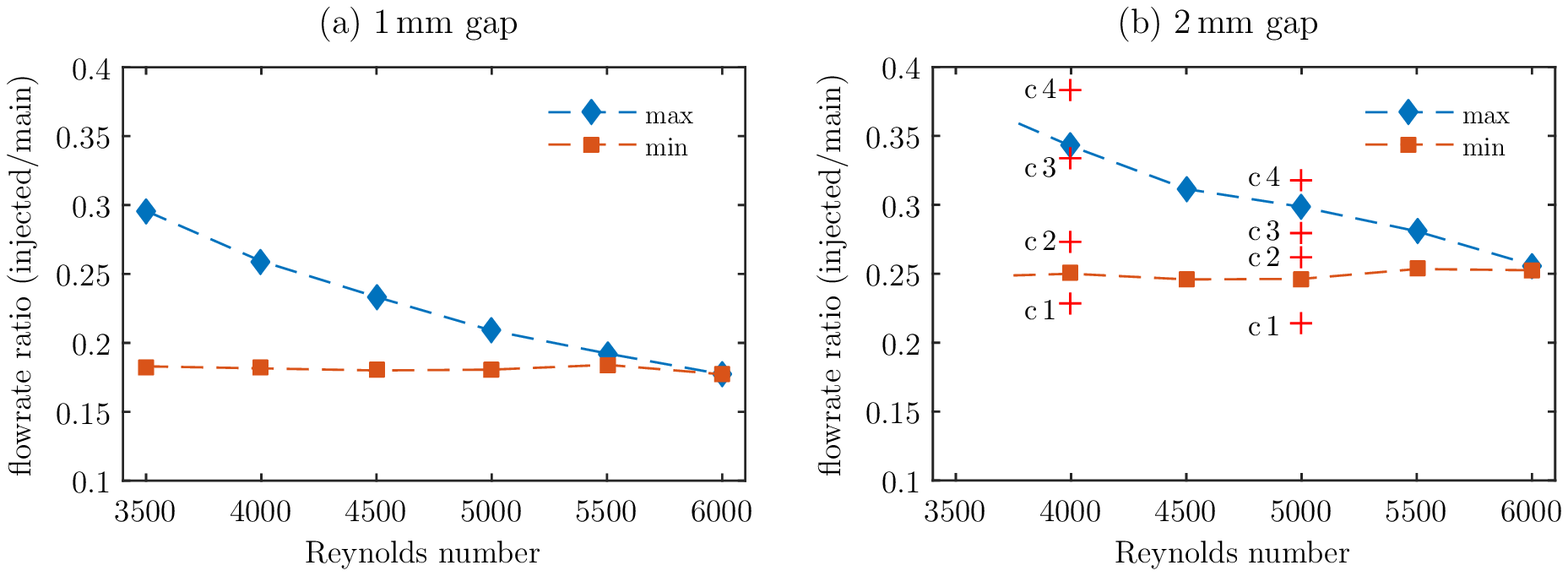}}
	\caption{\label{fig:nozzle1and2mm_inj-fr}Ratio of flow rate injected into the main pipe through the concentric gap. The gap width is a) 1\,mm and b) 2\,mm. For increasing Reynolds numbers max and min depict the maximum and minimum injection flow rates that cause total relaminarization. All flow rate ratios in between also lead to relaminarization. For the cases c1-c4 at $\Rey=4000$ and $\Rey=5000$ in (b) marked by a {\color{red}+} see the text.}
\end{figure}

\begin{figure}
	\centerline{\includegraphics[clip,width=0.98\textwidth,angle=0]{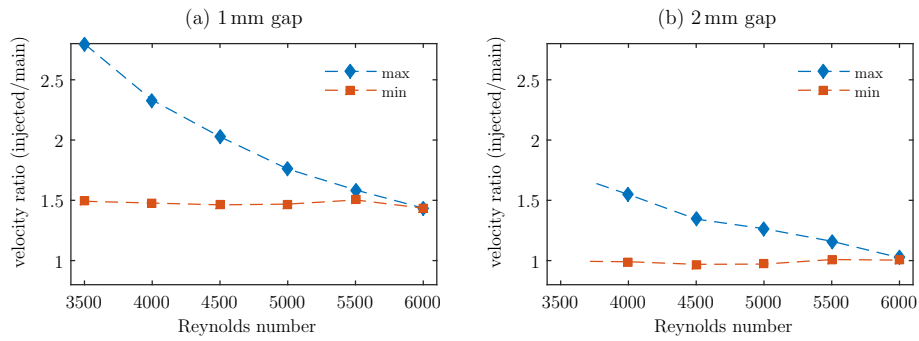}}
	\caption{\label{fig:nozzle1and2mm_vel-rat}Velocity ratios $U_{inj}/U_{n}$ of the injected flow to the respective main flow in the pipe at $z=0$ (in the plane of confluence). The (mean) velocities are calculated based on the measured flow rates as depicted in figure \ref{fig:nozzle1and2mm_inj-fr}.}
\end{figure}

With the 1\,mm-FMD and the 2\,mm-FMD, the devices which allow to inject fluid into the main pipe through an annular gap close to the pipe wall, the amount of injected fluid ($Q_{bp}$) and hence the level of acceleration close to the pipe wall can be controlled and continuously adjusted via a valve in the bypass. To investigate which flow rates and mean flow velocities of injected flow result in complete relaminarization, we increased the Reynolds number in increments of 500 and varied the flow rate through the bypass, i.e.\ the amount of fluid injected through the gap.

Up to $\Rey\cong6000$ we could find injection flow rates for both gap widths which cause a continuously laminar flow field at $z=150$ and the remainder of the pipe. I.e., when we measure the pressure drop sufficiently far downstream we find either turbulent or laminar pressure drop depending on the respective injection flow rate, nothing in between. Figure \ref{fig:nozzle1and2mm_inj-fr} shows the measured ratios $Q_{bp}/(Q_m-Q_{bp})$ of the flow rates of injected flow to the main flow for each gap width which lead to relaminarization. Figure \ref{fig:nozzle1and2mm_vel-rat} shows the respective velocity ratios $U_{inj}/U_n$ of injected flow to the main flow. In both figures max and min depict the maximum and minimum flow rate (velocity) ratios that would cause total relaminarization downstream. All flow rate (velocity) ratios in between also relaminarize the flow. While at lower Reynolds numbers a relatively broad range of flow rate ratios is suitable for relaminarization, the range quickly narrows down with increasing Reynolds number. For both devices it is rather the maximum of the flow rate ratio which decreases from initially $\sim30\%$ (1\,mm-FMD) and $\sim35\%$ (2\,mm-FMD), while the minimum amount of injected flow stays relatively constant at $18\%$ and $25\%$ for the 1\,mm-FMD and 2\,mm-FMD respectively. Concerning the velocity ratio this implies possible injection velocities in a broad range of 1.5-3 (1\,mm-FMD) and 1-1.5 (2\,mm-FMD) times higher than the mean flow velocity in the pipe at $\Rey=4000$. At $\Rey=6000$ the injection velocities necessary for total relaminarization narrow down to 1.5 and 1 as compared to the velocity of the main flow. For both gap widths the curves intersect slightly above $\Rey=6000$, indicating that for higher Reynolds numbers no complete relaminarization is possible with the present devices.  

Visual inspection of the flow field in the downstream vicinity ($0<z\lesssim50$) of the FMDs at Reynolds numbers and flow rate ratios between min and max show a scenario which is very similar to the one during relaminarization downstream the obstacle-FMD as described in section \ref{subsec:results-orifice}: the flow field in the direct downstream vicinity ($0< z\lesssim10$) of the device shows a turbulent flow and turbulence levels look comparable to the level upstream the device. However, for $\Rey\lesssim6000$ all visible perturbations in the flow field quickly decay downstream until the flow is clearly perceivable laminar at $z\approx30-50$. Furthermore, the flow stays unambiguously laminar for the remainder of the pipe.

If the ratio of injected fluid is just below the minimum or just above the maximum depicted in figure \ref{fig:nozzle1and2mm_inj-fr}, a transiently relaminarizing (laminarescent) part of the flow as in figure \ref{fig:relam-evo-pics} at $z=15$, where the turbulence intensity is obviously reduced by some amount in the downstream vicinity of the device, can be observed. I.e., incipient (transient) relaminarization at earlier stages is well observable, but finally the flow always returns to a fully turbulent state at around $z\approx30-50$.

As the scenario for injection flow rates between the indicated minimum and maximum and $\Rey\lesssim6000$ is so similar to the one downstream the obstacle-FMD at $\Rey=3800$ (see e.g.\ figure \ref{fig:relam-evo-pics} and \ref{fig:Orifice7holes_profile_evo}) and for flow rates closely below the minimum or above the maximum similar to to the one downstream the obstacle-FMD at $\Rey=4500$ respectively, no further pictures or measurements of the downstream evolution are shown for the 1\,mm-FMD and the 2\,mm-FMD.

\begin{figure}
	\centerline{\includegraphics[clip,width=0.94\textwidth,angle=0]{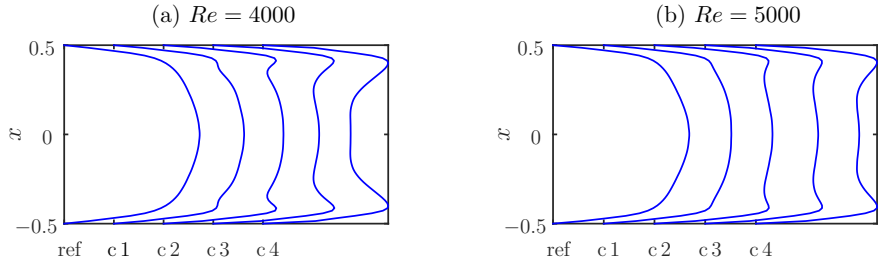}}
	\caption{\label{fig:inj_wall_2mm_profile_evo}Streamwise velocity profiles $(\bar w)$ measured at $z=2.5$ and (a) $\Rey=4000$ and (b) $\Rey=5000$. Cases $1-4$ depict different injection flow rates with the 2\,mm-FMD as indicated in figure \ref{fig:nozzle1and2mm_inj-fr} (b) by a {\color{red}+}. A measured profile of uncontrolled turbulent flow is shown for reference (ref).}
\end{figure}


Figure \ref{fig:inj_wall_2mm_profile_evo} presents PIV-measurements of 4 different injection flow rates with the 2\,mm-FMD at a single axial station, namely $z=2.5$, which is the closest distance to the device possible for our stereoscopic PIV-measurements. The four cases c1-c4 of different injection flow rates are indicated in figure \ref{fig:nozzle1and2mm_inj-fr} (b) by the {\color{red}+} symbols at $\Rey=4000$ and $\Rey=5000$. c1 represents a case where the flow rate of injected fluid is just beneath the minimum flow rate which is necessary for full relaminarization. c2 and c3 represent cases where full relaminarization is observed. At c4 the flow rate is already somewhat above the maximum, meaning that the controlled flow exhibits features of relaminarization but finally returns to a turbulent state downstream (similar to c1). The streamwise velocity profiles in figure \ref{fig:inj_wall_2mm_profile_evo} clearly show the increasing injection flow rate close to the wall. While for c1 only a minor hump is visible, for c2-c4 the peak close to the wall (stably at $x\cong0.41$ for all injection flow rates and both Reynolds numbers) is manifest.  

\begin{figure}
	\centerline{\includegraphics[clip,width=0.74\textwidth,angle=0]{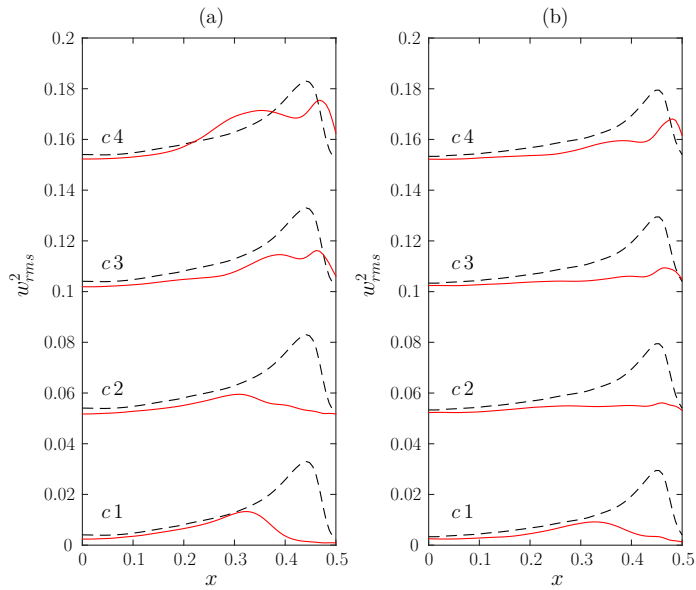}}
	\caption{\label{fig:wall_inj_2mm_wrms_evo}Streamwise velocity fluctuations ($w_{rms}^2$) measured at $z=2.5$ and (a) $\Rey=4000$ and (b) $\Rey=5000$. Cases c1$-$c4 depict different injection flow rates as indicated in figure \ref{fig:nozzle1and2mm_inj-fr} (b) by a {\color{red}+}. The dashed line is showing uncontrolled turbulent flow for reference.}
\end{figure}

The respective streamwise velocity fluctuations at $z=2.5$ are plotted in figure \ref{fig:wall_inj_2mm_wrms_evo}. For the injection case c1 the near wall peak is greatly reduced compared to the reference flow and moved from $x\cong0.44$ to $x\cong0.33$ for both Reynolds numbers. All streamwise fluctuations close to the wall have almost disappeared, and in the core region the rms level is already reduced too. However, the flow does not relaminarize in the long term. For injection cases c2 and c3 the reduction of the near wall peak is even more distinct, especially at $\Rey=5000$, where no significant peak is present at all. Also the reduction in the core region is more pronounced. Although for c3 the fluctuation level close to the wall is increased as compared to c2, noticeably especially for $\Rey=4000$, c2 and c3 show clearly reduced fluctuation levels and accordingly relaminarize completely downstream. For the injection case c4 it seems that, although the absolute level of the near wall peak is still reduced below the uncontrolled turbulent reference, a double humped, thickened shape of the peak can be observed. c4 eventually turns fully turbulent further downstream.

For Reynolds numbers $>6000$ we visually also observed a transiently relaminarizing state in the section $z\simeq6-30$ up to $\Rey\approx10000$, similar to the scenario described in section \ref{subsec:results-orifice} downstream the obstacle-FMD at $\Rey=4500$. The transient relaminarization can cause laminar-turbulent intermittency in the section $z\simeq30-100$, although not very pronounced. The higher the Reynolds number the higher the turbulent fraction, where laminar patches would quickly shrink downstream due to the faster propagation of the turbulent fronts \cite{Barkley2015}. 

\begin{figure}
	\centerline{\includegraphics[clip,width=0.94\textwidth,angle=0]{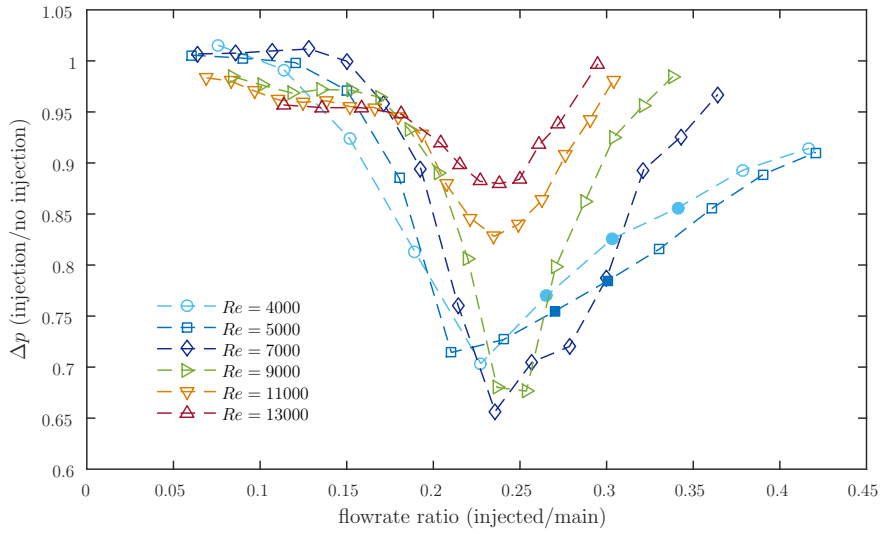}}
	\caption{\label{fig:rel_friction_rel_inj}Variation of the pressure drop $\Delta p$ (scaled with the pressure drop of the unforced reference flow) with different injection flow rates, measured in the downstream vicinity of the 2\,mm-FMD (first pressure tap at $z=6.2$, second at $z=29$). Filled symbols indicate full relaminarization further downstream.}
\end{figure}

To further investigate the evolution of the flow right downstream the injection device and detect temporally and spatially confined (partial) relaminarization we measured the mean streamwise pressure gradient between two pressure taps at $z=6$ and $z=29$ respectively (i.e.\ over a $23\,D$ stretch). The first pressure tap is supposed to be sufficiently far away from $z=0$ to avoid direct influence from the backward facing step of the injection device. At the same time the second pressure tap is supposed to be sufficiently close to the device to cover the whole area of transient relaminarization suggested by visual observations and figures \ref{fig:Orifice7holes_profile_evo} and \ref{fig:Orifice7holes_ti}. 

The measured variation of $\Delta p$ with the injection flow rate at several Reynolds numbers is plotted in figure \ref{fig:rel_friction_rel_inj}. For all cases depicted we find significant drag reduction at certain injection flow rates, and the trend is similar for increasing the injection from zero: above an injected flowrate of $\sim15\%$ a clear drag reduction can be observed, suggesting the onset of relaminarization for flowrate ratios roughly between $10-15\%$ for this specific device. The highest decrease in $\Delta p$ is found for flowrate ratios around $\sim20-25\%$.  With even higher injection flow rates $\Delta p$ returns towards the initial uncontrolled value. For $\Rey=4000$ and $\Rey=5000$ $\Delta p$ decreases by up to $\sim30\%$ for injection flow rates of $\sim20-25\%$. Interestingly, the maximum drag reduction over the $23\,D$ stretch downstream of the injection point is not achieved for those injection flow rates necessary for complete relaminarization further downstream (indicated by filled symbols, see also figure \ref{fig:nozzle1and2mm_inj-fr}), but for flow rates which are slightly below.

For $\Rey=7000$ and $\Rey=9000$ the drag reduction is even larger (down to 65\% of the uncontrolled flow), despite the fact that the flows return to turbulence further downstream. At $\Rey=11000$ and $\Rey=13000$ the decrease in $\Delta p$ is still significant (by 18\% and 12\% respectively), yet already much less. However, the measurements clearly indicate transient relaminarization also at these higher Reynolds numbers, i.e.\ temporary relaminarization in a spatially confined region right downstream the injection device.

\section{Discussion}\label{sec:discussion}

Our results demonstrate that by increasing the flow velocity close to the pipe wall in the proposed way turbulence can either be completely annihilated or at least temporally (in a spatially confined region downstream the control) weakened considerably. A chief precondition to the relaminarization process seems to be a steady, homogeneous local acceleration of the region of the viscous sublayer and the buffer layer above the level of uncontrolled flow, while the velocity is decreased accordingly in the log law region. Once this profile modification is realized, we observe an immediate collapse of turbulence production at the wall. Thereafter turbulence intensity decays exponentially with $z$. The most drastic reduction in turbulence levels is found in the region near the pipe wall. If the injection (forcing of the velocity profile) is too weak or too strong or at $\Rey\gtrsim6000$ several effects like remaining perturbations in the flow or secondary circulation caused by the injection trigger turbulence again and inevitably lead to a turbulent flow further downstream.

The results are in good accordance with the principal process of relaminarization in previous investigations concerning the natural decay of turbulence at low (subcritical) Reynolds numbers. E.g.\ \cite{Narayanan1968}, who investigated relaminarization by reducing the initial Reynolds number, found exponential decay of the turbulence intensity in the streamwise direction too. \cite{Sibulkin1962} also noted that the rate of decay of turbulence fluctuations appears to be more rapid near the wall and in the middle region of the pipe rather than at intermediate positions.

\cite{Pennell1972}, who similarly observed temporal relaminarization triggered by fluid injected through a porous-walled pipe at Reynolds numbers above criticality, noticed a thickening of the viscous and buffer layers due to the start of injection. This caused a significant yet transient reduction of the turbulence level. In the later stages of the re-transition to full turbulence (already at 6 and 10\,D downstream the injection) their rms-profiles exhibit a very similar double-humped shape as we observe in figure \ref{fig:wall_inj_2mm_wrms_evo} for case 4. These authors suggest that the double-humped rms-profile is a definite characteristic of the re-transition process. The hump nearest the wall is believed to result from the breakdown of the laminar layer. 


\begin{figure}
	\centerline{\includegraphics[clip,width=0.92\textwidth,angle=0]{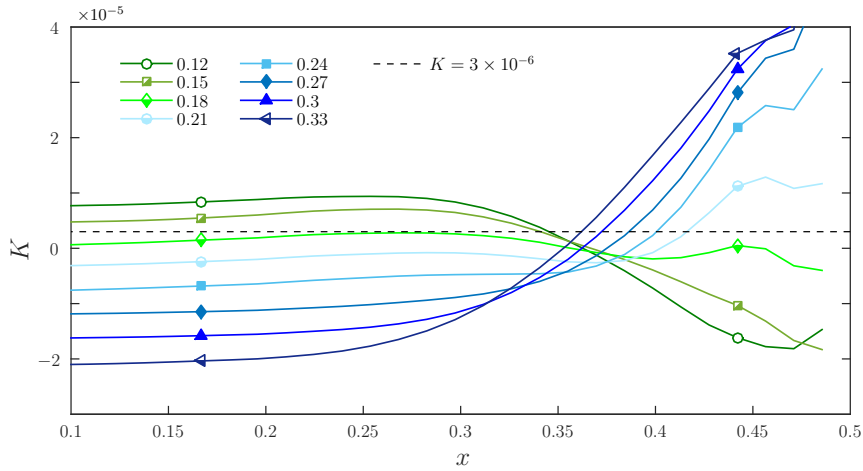}}
	\caption{\label{fig:k-re=5000}Local acceleration parameter $K$ for different injection flow rates by FMD\,2, measured at $\Rey=5000$ and $z=2.5$. Filled symbols indicate full relaminarization, half-filled symbols partial relaminarization further downstream.}
\end{figure}

Furthermore, the qualitative trend of our data is very similar to observations by e.g.\ \cite{Patel1968,Blackwelder1972,Narasimha1973,Spalart1986,Warnack1998,Ichimiya1998,Mukund2006} and \cite{Bourassa2009} concerning the effect of a favorable pressure gradient (FPG) on a turbulent boundary layer. Under the (accelerating) influence of a FPG the viscous sublayer of a turbulent boundary layer is known to increase in thickness and the velocity in the outer region decreases. Measurements and simulations of boundary layers under a FPG have been characterized to exhibit reversion from the turbulent to the laminar state. It is also the FPG and its effect on the wall region of the flow which has been found to be primarily responsible for departures from the inner law (and, by inference, for triggering relaminarization).

Various parameters have been suggested to quantify the acceleration level in spatially accelerating flows and define the onset of relaminarization, see \cite{Bourassa2009} for a compilation of 'acceleration driven laminarization parameters'. While there is not a consensus in criteria of laminarization, one of the most widely used is the acceleration parameter 
\begin{equation}\label{accpara}
K = \frac{\nu}{U_{\infty}^2} \,\frac{\textrm{d}U_{\infty}}{\textrm{d}z}
\end{equation} 
where z is the streamwise direction, $U_{\infty}$ is the free-stream velocity in this direction and $\nu$ the kinematic viscosity. Several authors have reported that turbulence is not sustained if $K$ is higher than a critical value in the range $2.5-3.6 \times 10^{-6}$ \cite{Launder1964,Moretti1965,Blackwelder1972,Spalart1986}. At the same time most researchers have also criticized the use of $K$ for being based on bulk flow parameters while the laminarization phenomenon is assumed to be a boundary layer event, necessarily having its basis in boundary layer considerations.

Based on eq.\ {\ref{accpara}} we calculated the acceleration parameter $K= \frac{\nu}{w^2} \,\frac{ \Delta w}{\Delta z}$ for different injection flow rates by FMD\,2 at $\Rey=5000$ with $\Delta z=2.5$, i.e.\ we compare the acceleration level at $z=2.5$ to the reference flow. Figure \ref{fig:k-re=5000} shows the result for selected flowrate ratios (0.12--0.33 in increments of 0.03) as a function of the radial direction. Furthermore, the aforementioned level of $K\approx3\times10^{-6}$ is indicated. Note that at arbitrarily small $z$ close to the backward-facing step of the injection device the values for $K$ in the near-wall region would be even higher. Figure \ref{fig:k-re=5000} suggests that the calculated values for $K$ very roughly match with the critical values proposed in the literature, in particular when taking into account the findings provided by figure \ref{fig:rel_friction_rel_inj}, indicating the onset of relaminarization for flowrate ratios between $10-15\%$ for FMD\,2. It would need measurements much closer to $z=0$ for a more accurate assessment. 

However, the major drawback of $K$ is in any case the empirical nature and the lack of a mechanistic explanation. \cite{Warnack1998} also pointed out that no single criterion could be used to predict the beginning or end of the breakdown of the law of the wall. They found that for flow with low acceleration where laminarization did not occur, there still was a breakdown of the law of the wall and a slight reduction in turbulent intensities. Relaminarization in boundary layers was observed as a gradual change of the turbulence properties and not catastrophic. Retransition, however, is a fast process due to the remaining turbulence structure and may be compared with bypass transition \cite{Warnack1998}.

An interesting approach to the question how the specific conditions can cause the flow to laminarize on a local scale is provided by \cite{Jimenez1999}. They have shown not only that the near-wall turbulence regeneration cycle is autonomous, i.e.\ self-sustaining independently of whether or not fluctuations are present in the region $y^{+}>60$, but also that it can be interrupted numerically at various places, leading to the decay of turbulence and to eventual laminarization. Using a filter to make the streamwise velocity (more) uniform they identified the minimum streak length needed to sustain the cycle to be between 300 and 400 wall units. Their proposed control strategy is hence to weaken or decorrelate the streaks in the region below $y^{+}=60$ and above $y^{+}=20$.

\begin{figure}
	\centerline{\includegraphics[clip,width=0.88\textwidth,angle=0]{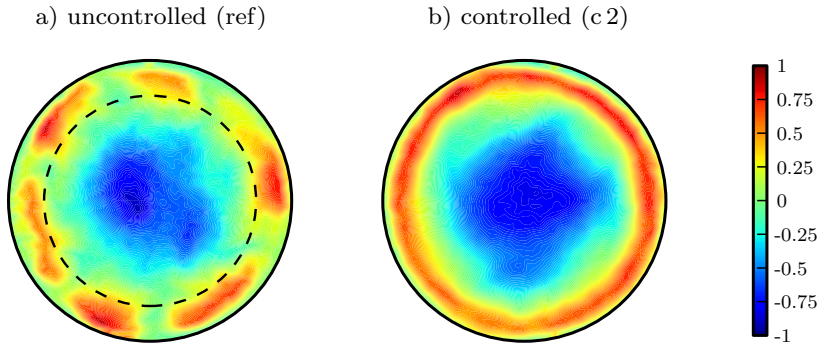}}
	\caption{\label{fig:streakcontours}(Colour online) Contours of the (instantaneous) streamwise velocity $w$ in the cross-section of the pipe. The laminar flow field has been subtracted to emphasize near wall streaks (color bar expressed in units of the bulk velocity). a) is the unmodified turbulent reference flow and b) depicts case c\,2 with an injection flow rate as indicated in figure \ref{fig:nozzle1and2mm_inj-fr} (b) by a {\color{red}+} ($\Rey=4000$, $z=2.5$). $y^{+}\approx60$ is indicated by the dashed circle.}
\end{figure}

Instead of using a numerical filter on the near-wall region, we locally replace and substitute the flow by injection through the gap. The injected flow and the resulting, modified flow in the near-wall region is supposedly more homogeneous and 'streak-free'. In support of this, figure \ref{fig:streakcontours} (a) shows several distinct (high-speed) streaks in the near wall region (the region roughly below $y^{+}=60$ marked by the dashed circle) of an instantaneous snapshot of turbulent flow at $\Rey=4000$. However, in (b), depicting the case c\,2 of relatively small injection just sufficient for full relaminarization, the streaky structures in the near-wall region have disappeared.


\cite{Kuehnen2018} designate the relaminarization observed due to a modified profile also to a weakening of the near-wall cycle, yet argue from a slightly different perspective focusing on the efficiency of the 'lift up mechanism' \cite{Brandt2014}. To obtain a measure for the amplification mechanism of the regeneration cycle they consider the linearized Navier-Stokes equations and perform a transient growth (TG) analysis (following the algorithm given by \cite{Meseguer2003}). The velocity profiles of all successfully modified, i.e.\ relaminarizing flows considered by \cite{Kuehnen2018} are shown to exhibit a substantially reduced transient growth. Generally the flatter the velocity profile the more the streak vortex interaction is suppressed and in the limiting case of a uniformly flat profile the lift up mechanism breaks down entirely. 

\begin{figure}
	\centerline{\includegraphics[clip,width=0.92\textwidth,angle=0]{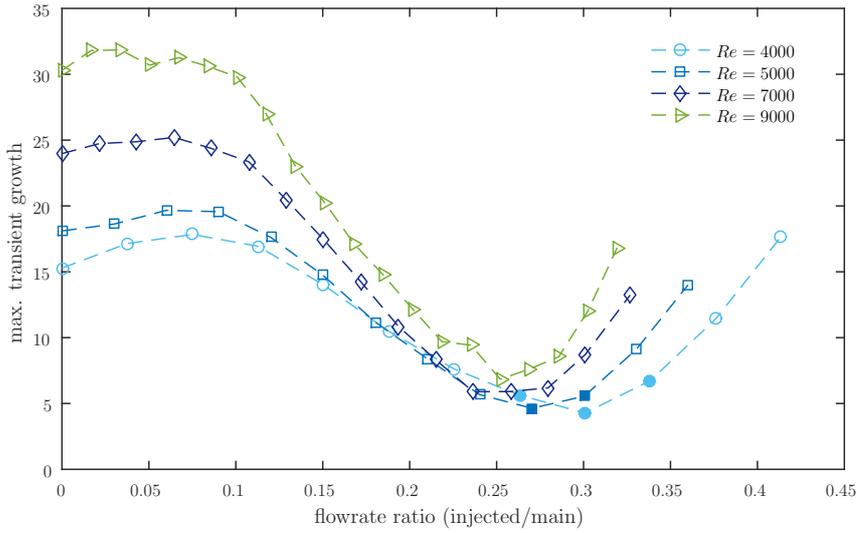}}
	\caption{\label{fig:transientgrowth}Variation of the transient growth with different injection flow rates, measured right behind (at $z=2.5$) the 2\,mm-FMD. Filled symbols indicate full relaminarization further downstream.}
\end{figure}

We applied the same procedure here to (azimuthally and temporally averaged) velocity profiles measured at $z=2.5$ downstream of the 2\,mm-FMD for different injection flow rates. Assuming that the profile is fixed under the influence of the perturbation we conducted a TG analysis around the modified profiles. As shown in figure \ref{fig:transientgrowth} the profiles indeed show a considerably decreasing TG with increasing injection flowrate suggesting that vortices are less efficient in producing streaks. Very similar to the trend of the pressure drop measured in the downstream vicinity of the device (see fig.\ \ref{fig:rel_friction_rel_inj}) the TG exhibits a clear minimum amplification at an optimal flowrate ratio before it starts to rise again towards the value of unmodified flow. Interestingly, the optimal is found at slightly higher injection flowrates than those for the smallest pressure drop in fig.\ \ref{fig:rel_friction_rel_inj}. In any case, the flowrate ratios with the least amplified perturbations are exactly those which relaminarize downstream at $\Rey=4000$ and $\Rey=5000$ (indicated with filled symbols). The data suggest that the turbulence regeneration cycle is weakened indeed and breaks down due to a reduced efficiency of the lift up process.

The present study elucidates the general process of relaminarization caused by a modified streamwise velocity profile by simple means. Both the stationary obstacle-FMD and the injection nozzles (1\,mm-FMD and 2\,mm-FMD) can cause total relaminarization of the flow downstream and by that reduce skin friction by large amounts. As laminar pipe flow is linearly stable \cite{Drazin1981} it will stay laminar for the remainder of a smooth straight pipe. Although both devices cause an initial cost either by the pipe blockage or by the required injection for full relaminarization, it is only a question of how long the unperturbed (straight and smooth) downstream pipe section is until a net energy gain can be realized. As both devices are used solely for demonstration and proof-of-principle purposes we do not consider the possible net energy gain achievable with the present devices.

\section{Summary and conclusion}\label{sec:summary}

Initially, relaminarization due to flow acceleration close to the wall may seem counterintuitive. However, in this work (see also \cite{Kuehnen2018}) we demonstrate that fully turbulent flows in pipes can be completely relaminarized by rather simple means. An open-loop forcing of the streamwise velocity profile, either actively or passively, is sufficient to trigger turbulence breakdown. The present experiments demonstrate that the onset of relaminarization in a fully developed turbulent pipe flow occurs as a direct result of a particular shear-stress distribution in the wall-region. The underlying physical mechanism of relaminarization is attributed to a weakening of the near-wall turbulence production cycle.

Two different devices, a stationary obstacle (inset) and a device to inject additional fluid through an annular gap close to the wall, are used to control the flow. Both devices modify the streamwise velocity profile such that the flow in the center of the pipe is decelerated and the flow in the near wall region is accelerated. Visualization, pressure drop measurements and stereoscopic PIV measurements have been employed to determine both practice-oriented and fundamental information for relaminarizing turbulent flows in a circular pipe in the presence of devices which accelerate the near wall region. Downstream of the devices a fully developed laminar flow is established. The fluctuations are greatest just beyond the devices and die away with increasing downstream distance.

High amounts of energy (pumping power due to frictional drag) can be saved if the flow is laminar instead of turbulent. At $\Rey=6000$, the highest Reynolds number were we achieve full relaminarization with the present devices, the pressure drop in the downstream distance is reduced by a factor of $3.4$ due to relaminarization. At e.g.\ $\Rey=13\,000$, where we find transient (temporary) relaminarization in a spatially confined region right downstream the devices, the drag reduction is still higher than 10\%.

Future research should focus on the possible net energy gain and an optimization of the specific design in terms of pressure loss, such that an energetic break even point can be reached in a close downstream distance of the devices. Furthermore, a smarter design may allow relaminarization at much higher Reynolds numbers. In order to establish a possible cost saving potential of the presented control technique it is also necessary to determine over which distances the relaminarized flow persists under less perfect pipeline conditions.

\begin{acknowledgements}
The project was partially funded by the European Research Council under the European Union’s Seventh Framework Programme (FP/2007-2013)/ERC grant agreement 306589. The authors declare that they have no conflict of interest. We thank George H.\ Choueiri for friendly help and useful discussions. We thank M.\ Schwegel for a Matlab-code to post process experimental data.

\end{acknowledgements}

\bibliographystyle{abbrv}
\bibliography{Literatur_JK}   

\end{document}